\documentclass[12pt]{article}

\usepackage[english]{babel}

\usepackage[a4paper,top=2cm,bottom=2cm,left=3cm,right=3cm,marginparwidth=1.75cm]{geometry}

\usepackage{amsmath}
\usepackage{graphicx}
\usepackage{authblk}
\usepackage[colorlinks=true, allcolors=blue]{hyperref}
\usepackage{epsfig}
\usepackage{amssymb}
\usepackage{xcolor}
\usepackage{subfigure}
\usepackage{bm}
\usepackage{dsfont}
\usepackage{hyperref}
\usepackage{todonotes}
\usepackage{ulem}
\usepackage{physics}
\usepackage{easyReview}
\usepackage{cite}
\title{Squeezing of a coupled state of two spin-1 systems}
\author{Akhilesh K. S.}
\affil{Department of Physical Sciences\\
Indian Institute of Science Education \& Research (IISER) Mohali\\
Sector 81 SAS Nagar, Manauli, 140306, Punjab, India}
\date{}
\begin{document}
\maketitle

\begin{abstract}
\noindent The nature of spin squeezing has been studied earlier for a coupled state of two spinors by Usha Devi et. al. (J. Phys. A: Math. Gen. {\bf 36} 5333 (2003)).  In this paper, we extend this study to a coupled state of two spin-1 systems. Here,  we show that the direct product of two spin-1 states exhibits squeezing unlike the direct product of two spinors.  We investigate the nature of squeezing for non-separable coupled state of two spin-1 systems and identify the parameters which are responsible for the existence of squeezing. We also provide ways to generate them by external interactions.
\end{abstract}

\section{Introduction}\label{sec1}

 Spin squeezing has wide applications in the field of high precision spectroscopy\cite{caves,yurke,kit1}, quantum enhanced metrology\cite{pezze}, detection of quantum chaos, and quantum phase transitions\cite{sachdev,haake}. Importantly, it is treated as a criterion for detection of entanglement in multiqubit systems\cite{julsgaard,riedel}. It refers to reduction of the variance of a spin component in the plane perpendicular to the mean spin direction below a certain limit.  Kitegawa and Ueda~\cite{kit2} have made a detailed study of spin squeezing and proposed a coordinate independent definition of spin squeezing. They have shown that the existence of squeezing is due to the consequence of quantum spin-spin correlations that exist among 2$s$ spinor states which together constitute the spin state. For coherent spin states, the variance of spin components are equally distributed in the plane perpendicular to the mean spin direction. However,  for spin squeezed states, the value of variance in the plane perpendicular to mean spin direction is below the limit set by the coherent states.  Later, Puri \cite{puri} has proposed a more stringent criterion for spin squeezing. In this criterion, the limit is set by the expectation value of the spin component along the mean spin direction. We make use of this criterion in our paper. 
  	
  	The study of spin squeezing can also be extended to  bipartite systems. A bipartite state can  be of two forms: separable or entangled. The squeezing in the bipartite system depends on the self and the mutual spin-spin correlations that exist in the system. The aspect of squeezing in coupled state of two spinors is studied by Usha Devi et. al.~\cite{usha1}.   We extend this study to coupled state of two spin-1 systems. Spin-1 systems are experimentally available. Recently, deuterium
  	spin, which is spin-1, in a deuterated chloroform molecule oriented in a
  	lyotropic liquid crystal used for implementing qutrit gates~\cite{dogra}. Two Rydberg atoms behave as a spin-1 system used for implementing shortcut to adiabaticity~\cite{rydberg,erik}. Considering two two-level atomic states, Barnette and Duperturs have constructed squeezed atomic states in anology with the multimode squeezed states and thermofield states of the radiation field~\cite{barnett}. This work will give ways of constructing coupled squeezed states of two spin-1 systems. 

In this work, we study the nature of squeezing for a pure coupled state of two spin-1 systems and suggest ways to generate them. This paper is organised as follows. In section \ref{sec:coupled-systems}, we discuss coupled systems. In section \ref{sec:squeezing-coupled}, we briefly explain the criteria for spin squeezing in coupled systems. In section \ref{sec:squeezing-two-spin-1}, we study the nature of squeezing by considering both separable and entangled coupled state in the context of two spin-1 subsystems. In section \ref{sec:generation}, we suggest ways of generating squeezed and non-squeezed states. Section \ref{sec:conclusion} contains the conclusion.


\section{Coupled systems}\label{sec:coupled-systems}

 A bipartite system has two subsystems. Let $s_1$ and $s_2$ be their spins. Then, the general coupled state of these two subsystems can be written as
\begin{align}
\ket{\psi_{12}}=\sum_{ij}c_{ij}\ket{\phi_i}\otimes \ket{\chi_j},
\end{align}
where $\{\ket{\phi_i};i=-s_1,\cdots, s_1\}$ and $\{\ket{\chi_j};j=-s_2,\cdots,s_2\}$ are the angular momentum bases of the subsystems. These bases are chosen with respect to some axes of quantization. Let $\hat{Q}_1$ and $\hat{Q}_2$ be their axes of quantization. The coupled states which are the eigenstates of the four commuting operators $\hat{S}_1.\hat{Q}_1,\hat{S}_2.\hat{Q}_2,\hat{S}_1^2,\hat{S}^2_2$ are called as oriented states~\cite{mallesh}. More precisely,
\begin{align}
(\vec{S}_1.\hat{Q}_1)(\vec{S}_2.\hat{Q}_2)\ket{\psi_{12}}=m_1 m_2 \ket{\psi_{12}}.
\end{align}
  It has been shown that the oriented states do not exhibit squeezing~\cite{mallesh}. All other states are termed as non-oriented states which may exhibit squeezing behaviour. For $s_1=s_2=1/2$, all the direct product couple states are oriented states with respect to appropriate choices of axes of quantization. Therefore, only entangled coupled states exhibit squeezing~\cite{usha1}. But for $s_1,s_2>1/2$, not all the direct product states are the eigenstates of the four commuting operators $\hat{S}_1.\hat{Q}_1,\hat{S}_2.\hat{Q}_2,\hat{S}_1^2,\hat{S}^2_2$. So, some direct product states may show the presence of squeezing. Because we know that any spin $s$ state can be realized by $2s$ number of spin half states, the correlation among $2s$ spin half states can lead to existence of spin squeezing. In the next subsequent section, we study the nature of squeezing for a coupled state of two spin-1 systems. First, let us study the criterion for squeezing in the case of coupled systems.

\section{Squeezing of coupled systems}\label{sec:squeezing-coupled}

Spin coherent state is a simultaneous eigenstate of the operators $\hat{S}^2$ and $\hat{S}.\hat{n}$, where $\hat{n}$ is the axis of quantization, with eigenvalues $s(s+1)$ and $\pm s$. For the spin coherent states, the Heisenberg uncertainty relation takes the form:
\begin{align}
\Delta (\hat{S}.\hat{n}_{\perp})^2 \Delta (\hat{S}.\hat{n}_{\perp'})^2=\frac{s^2}{4}
\end{align}
with $\Delta (\hat{S}.\hat{n}_{\perp})^2 =\Delta (\hat{S}.\hat{n}_{\perp'})^2=s/2 $. Here $\hat{n}_{\perp}$ and $\hat{n}_{\perp'}$ are any two perpendicular directions in a plane perpendicular to the mean spin direction $\hat{n}$.

 According to Kitegawa and Ueda~\cite{kit2}, spin squeezed states are the states which have variance less than $s/2$ in a direction perpendicular to the mean spin direction. 
A state satisfying
\begin{align}
\Delta(\vec{S}.\hat{n}_\perp)< \frac{s}{2},
\end{align}
where $\Delta(\vec{S}.\hat{n}_{\perp})^2=\expval{(\vec{S}.\hat{n}_{\perp})^2}-\expval{\vec{S}.\hat{n}_{\perp}}^2$, $\vec{S}.\hat{n}_{\perp}$ is the spin operator in a direction perpendicular to the mean spin direction, is said to be spin squeezed. Later Puri \cite{puri} has proposed a more stringent condition given by
\begin{align}
\Delta(\vec{S}.\hat{n}_\perp)< \frac{\abs{\expval{\vec{S}.\hat{n}}}}{2}.
\end{align}
Here, the limit is set by the expectation value of the spin operator along the mean spin direction.

In the case of coupled systems, a coupled state with mean spin directions $\hat{n}_1$ and $\hat{n}_2$ is said to be squeezed if 
\begin{align}
\Delta(\hat{S}_1.\hat{n}_{1\perp}+\hat{S}.\hat{n}_{2\perp})^2<\frac{\abs{\expval{\vec{S}_1.\hat{n}_{1}}}+\abs{\expval{\vec{S}_2.\hat{n}_2}}}{2},
\end{align}
where $\hat{n}_{1\perp}$ and $\hat{n}_{2\perp}$ are the two directions in a plane pependicular to mean spin directions $\hat{n}_1$ and $\hat{n}_2$ respectively\cite{usha1}.
The above condition reduces to the form

\begin{align}
\Delta(\vec{S}_1.\hat{n}_{1\perp})^2+\Delta(\vec{S}_2.\hat{n}_{2\perp})^2+2\expval{\vec{S}_1.\hat{n}_{1\perp}\otimes\vec{S}_2.\hat{n}_{2\perp}} < \frac{\abs{\expval{\vec{S}_1.\hat{n}_{1}}}+\abs{\expval{\vec{S}_2.\hat{n}_2}}}{2}.
\end{align}
Therefore, the squeezing parameter can defined to be
\begin{align}
\label{squeezing-parameter}
\xi = \frac{2\Delta(\vec{S}_1.\hat{n}_{1\perp})^2+2\Delta(\vec{S}_2.\hat{n}_{2\perp})^2+4\expval{\vec{S}_1.\hat{n}_{1\perp}\otimes\vec{S}_2.\hat{n}_{2\perp}}}{\abs{\expval{\vec{S}_1.\hat{n}_{1}}}+\abs{\expval{\vec{S}_2.\hat{n}_2}}}.
\end{align}
The states which satisfy $\xi<1$ are said to be squeezed states. Employing this parameter we study the squeezing behavior of  coupled states of two spin-1 systems. We consider both direct product states and entangled states in the context of two spin-1 systems and analyse their behaviour of squeezing. 

\section{Squeezing of a coupled state of two spin-1 systems}\label{sec:squeezing-two-spin-1}
This section has two parts: first part deals with the study of squeezing in direct product states of two spin-1 states, the  second part deals with the states which cannot be written as direct product of two spin-1 states. 
\subsection{Squeezing in direct product state of two spin-1 states}
In this section, we consider direct product of two spin-1 states and study the nature of squeezing in them. A general coupled state of two spin-1 systems can be written as
\begin{align}
\ket{\psi}=\begin{pmatrix}
c_{11} & c_{12} & c_{13} & c_{21} & c_{22} & c_{23} & c_{31} & c_{32} & c_{33}
\end{pmatrix}^T.
\end{align}
This state is said to be direct product of two spin-1 states if 
\begin{align}
c_{ij}c_{kl}=c_{il}c_{kj}
\end{align}
for all $i,j,k,l$. A general direct product state is written as
\begin{align}
\label{product}
\ket{\psi}=\ket{\psi}_1\otimes\ket{\psi}_2,
\end{align}
where $\ket{\psi}_1$ and $\ket{\psi}_2$ are the two spin-1 states.

\begin{figure}
 	\begin{center}
 	\includegraphics[width=3.5in]{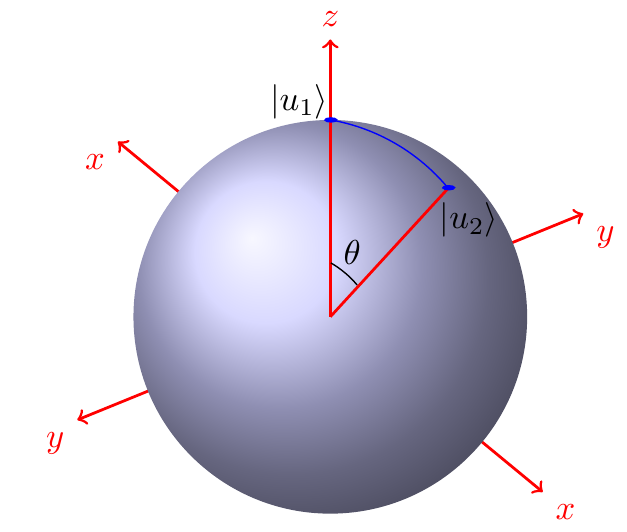}
 	\caption{\label{Bloch-sphere}Spin-1 state is represented by two spinors on the Bloch sphere. Choosing a coordinate system such that one spinor is along the $z$-axis and the other one is fixed in the $x$-$z$ plane.}
 \end{center}
 \end{figure}

 According to Schwinger construction~\cite{schwinger}, every spin $s$ state can be written as $2s$ number of spinors. Thus, one can write each spin-1 state as symmetric combination of two spinors. Therefore, we write each $\ket{\psi}_{i}$ as
 \begin{align}
 \ket{\psi}_i= \frac{1}{\mathcal{N}_i}\Big[\ket{u_1}_i\ket{u_2}_i+\ket{u_2}_i\ket{u_1}_i\Big];\,\,i=1,2,
 \end{align}
 where $\ket{u_1}$ and $\ket{u_2}$ are the spinors represented on the Bloch sphere.
 To study squeezing, we need to find mean spin directions. For that, first we need to fix coordinate axes for each subsystems. For each spin-1 subsystem, without loss of generality, one can choose a coordinate system such that the $z$-axis is along one of the spinors and the other spinor is fixed in the $x$-$z$ plane(see fig.(\ref{Bloch-sphere})). This leads to 
 \begin{align}
 \label{spinors}
 \ket{u_1}_i=\begin{pmatrix}
 1\\
 0
 \end{pmatrix},\,\,\,\,\,\ket{u_2}_i=\begin{pmatrix}
 \cos(\theta_i/2)\\
 \sin(\theta_i/2)
 \end{pmatrix}.
 \end{align}
  Then, 
 \begin{align}
 \label{product-two}
 \ket{\psi}_{i}=\frac{1}{\sqrt{3+\cos\theta_i}}\begin{pmatrix}
 2\cos(\theta_i/2)\\
 \sqrt{2}\sin(\theta_i/2)\\
 0
 \end{pmatrix}.
 \end{align}
 
 Now let us find mean spin directions. We need expectation values of spin operators along each axes of the coordinate system. 
 The spin-1 operators along $x,y, z$ directions are given by
 \begin{align}
 \hat{S}_x=\frac{1}{\sqrt{2}}\begin{pmatrix}
 0 & 1 & 0\\
 1 & 0 & 1\\
 0 & 1 & 0
 \end{pmatrix}, \hat{S}_y=\frac{1}{\sqrt{2}}\begin{pmatrix}
 0 & -i & 0\\
 i & 0 & -i\\
 0 & i & 0
 \end{pmatrix}, \hat{S}_z=\begin{pmatrix}
 1 & 0 & 0\\
 0 & 0 & 0\\
 0 & 0 & -1
 \end{pmatrix}.
 \end{align}
 The mean spin direction for each spin-1 subsystem is expressed as
 \begin{align}
 \hat{n}_i=\left(\frac{\expval{\hat{S}_{ix}}}{\expval{\bf{\hat{S}}}},\frac{\expval{\hat{S}_{iy}}}{\expval{\bf{\hat{S}}}},\frac{\expval{\hat{S}_{iz}}}{\expval{\bf{\hat{S}}}}\right)\,\,i=1,2;
 \end{align}
 where $\expval{\hat{S}_{1x}}=\expval{\hat{S}_{x}\otimes I},\expval{\hat{S}_{2x}}=\expval{I \otimes \hat{S}_{x}} $ similarly for $y$ and $z$,
 \begin{align}
 \expval{\bf{\hat{S}}}=\sqrt{\expval{\hat{S}_{ix}}^2+\expval{\hat{S}_{iy}}^2+\expval{\hat{S}_{iz}}^2}.
 \end{align}
 For the considered state(see eq.(\ref{product}) and eq.(\ref{product-two})), we get
 \begin{align}
  \hat{n}_i =\left(\frac{\sin\theta_i}{\sqrt{2(1+\cos\theta_i)}},0,\sqrt{\frac{(1+\cos\theta_i)}{2}}\right).
 \end{align}
 The two other mutually perpendicular directions are found to be
 \begin{align}
 \hat{n}_{i\perp}&=\left(\sqrt{\frac{(1+\cos\theta_i)}{2}},0,\frac{-\sin\theta_i}{\sqrt{2(1+\cos\theta_i)}} \right), \\
 \hat{n}_{i\perp'}&=(0,1,0).
 \end{align}
 Using eq.(\ref{squeezing-parameter}), the squeezing parameter is obtained as
 \begin{align}
 \label{squeezing-parameter-1}
 \xi =\dfrac{\Bigg(\dfrac{(1+\cos\theta_1)}{3+\cos\theta_1}+\dfrac{(1+\cos\theta_2)}{3+\cos\theta_2} \Bigg)}{\abs{\dfrac{\sqrt{2(1+\cos\theta_1)}}{3+\cos\theta_1}}+\abs{\dfrac{\sqrt{2(1+\cos\theta_2)}}{3+\cos\theta_2}}}.
 \end{align}
 \begin{figure}
 	\begin{center}
 		\includegraphics[width=4.5in]{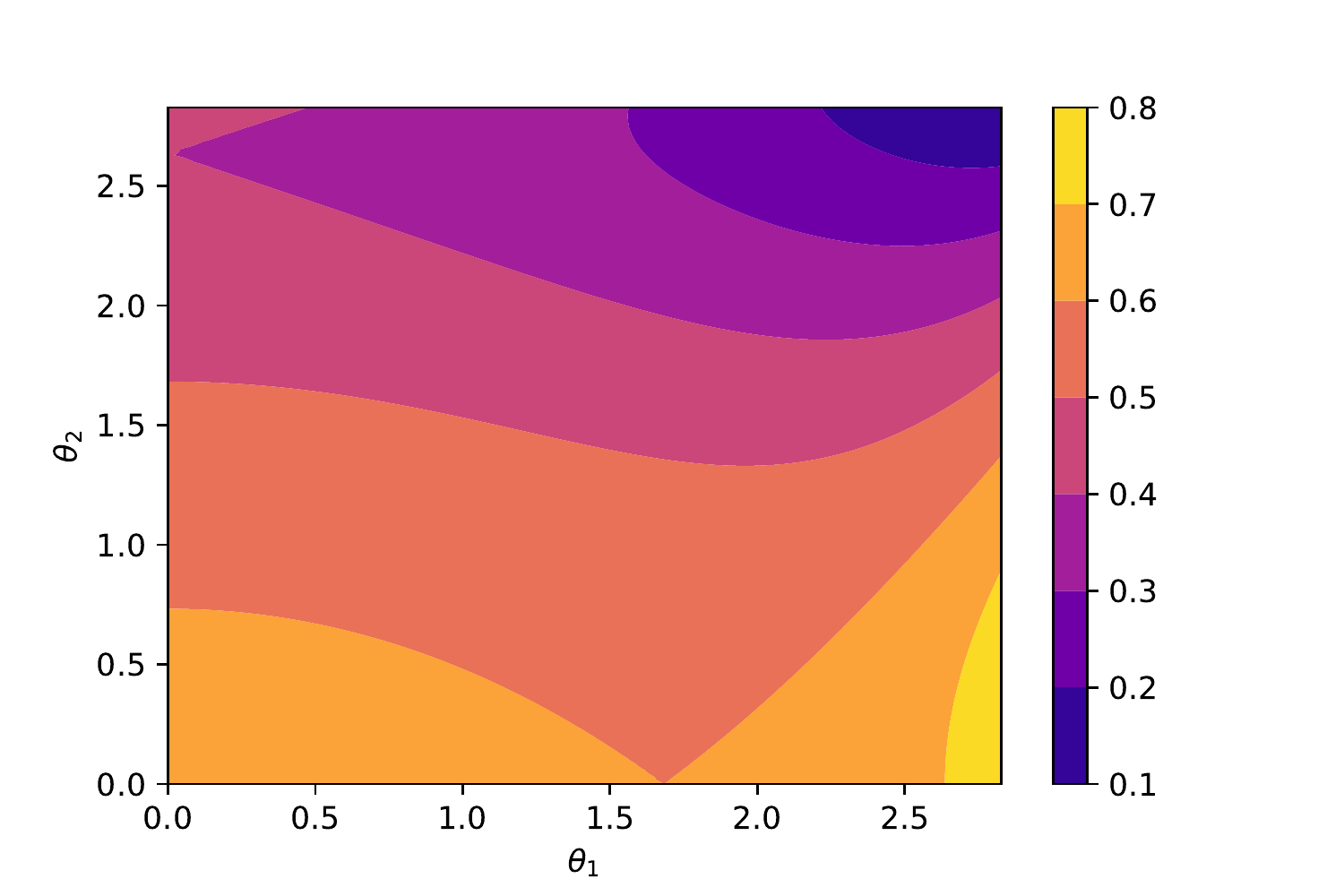}
 		\caption{\label{squeezing-1}Variation of squeezing parameter $\xi$  given in eq.(\ref{squeezing-parameter-1}) with respect to $\theta_1$ and $\theta_2$.}
 	\end{center}
 \end{figure} 
This parameter decides whether the direct product state is squeezed or not. We have plotted the squeezing parameter for various values of $\theta_i$'s (see fig.(\ref{squeezing-1})). From fig.(\ref{squeezing-1}), we can see that the direct product of spin-1 states exhibit squeezing unlike the direct product state of two spinors. The nature of squeezing is observed in the whole range of $\theta_i$'s. The amount of squeezing has reached around 0.1.
 
 Let us consider a coupled state which is a direct product of a coherent state and a squeezed state, i.e., 
 \begin{align}
 \begin{pmatrix}
 1\\
 0\\
 0
 \end{pmatrix} \otimes\frac{1}{\sqrt{3+\cos\theta}}\begin{pmatrix}
 2\cos(\theta/2)\\
 \sqrt{2}\sin(\theta/2)\\
 0
 \end{pmatrix}.
 \end{align}
 For this state, the squeezing parameter turns out to be
 \begin{align}
 \label{squeezing-parameter-3}
 \xi = \dfrac{1+\dfrac{(1+\cos\theta)}{3+\cos\theta} }{1+\abs{\dfrac{\sqrt{2(1+\cos\theta)}}{3+\cos\theta}}}.
 \end{align}
 
\begin{figure}
 	\begin{center}
 		\includegraphics[width=4in]{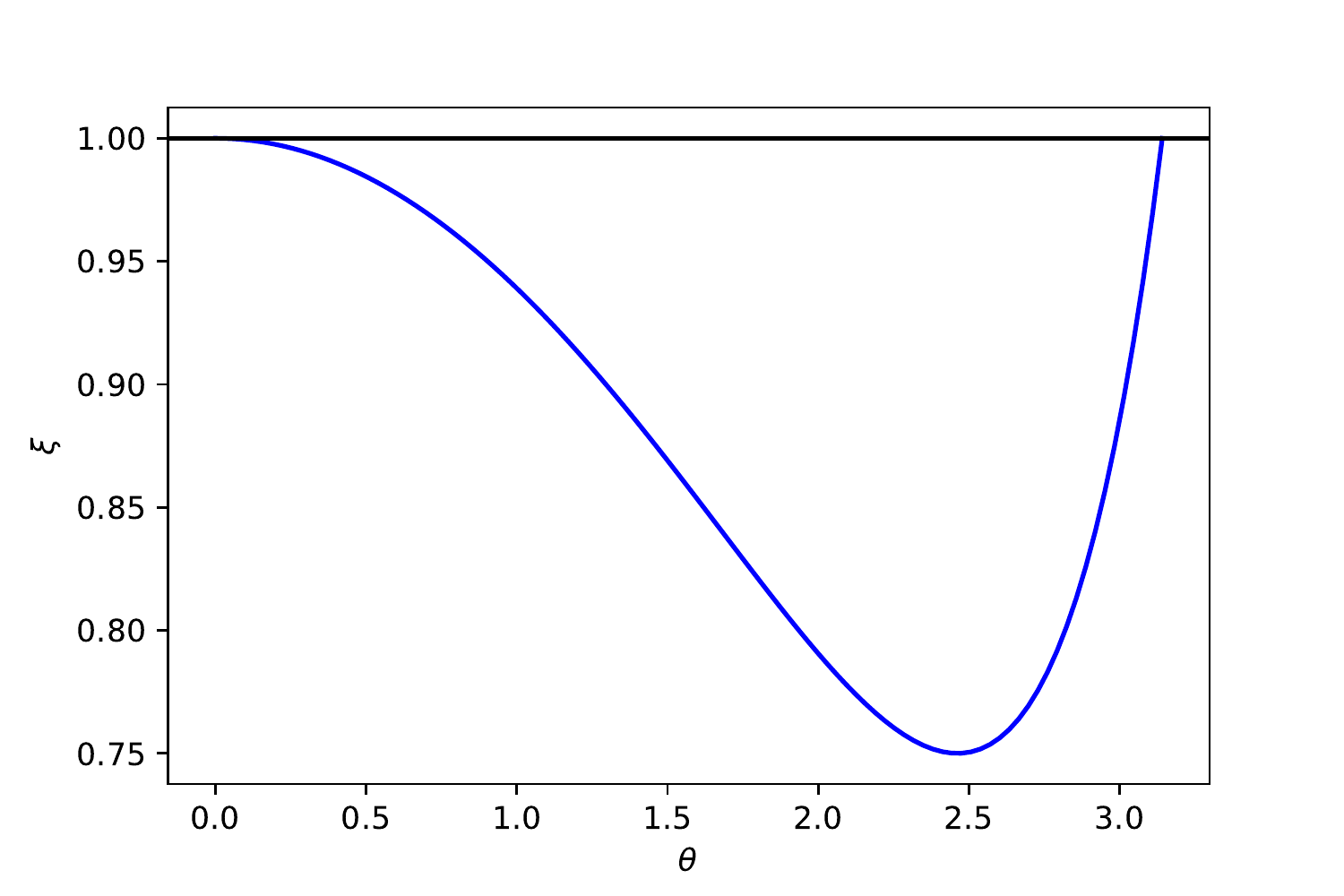}
 		\caption{\label{figure3}Variation of squeezing parameter $\xi$ given in eq.(\ref{squeezing-parameter-3}) with respect to $\theta$.}
 	\end{center}
 \end{figure}

 By plotting the squeezing parameter (see fig.(\ref{figure3})), we observe that the range of squeezing is not affected by replacing squeezed state by a coherent state but the amount of squeezing has been decreased to 0.75. This shows that the coupled state with direct product of two squeezed states are more suitable for high precision measurements than the coupled state consisting only one squeezed state. 
       
\subsection{Squeezing in entangled state of two spin-1 systems}
Here, we study the behavior of squeezing for the coupled states which cannot be written as direct product of two spin-1 states. The general state of coupled two  spin-1 systems is written as
\begin{align}
\ket{\psi}=\begin{pmatrix}
c_{11} & c_{12} & c_{13} & c_{21} & c_{22} & c_{23} & c_{31} & c_{32} & c_{33}
\end{pmatrix}^T.
\end{align}
It cannot be written as direct product of two spin-1 states if 
\begin{align}
c_{ij}c_{kl}\ne c_{il}c_{kj}
\end{align}
for at least one of $i,j,k$ and $l$ values. Let us explore the nature of squeezing for such states. To study squeezing, we need to choose coordinate systems. Without loss of generality, one can choose the $z$-axes along the mean spin directions. This requires
\begin{align}
\label{condition-for-z-axes}
\expval{S_x\otimes I}=\expval{I\otimes S_x}=0,\\
\label{condition-for-z-axes2}
\expval{S_y\otimes I}=\expval{I\otimes S_y}=0.
\end{align}
Because of this choice, we can write $c_{23}$ and $c_{21}$ as
\begin{align}
\label{conditions-msd-along-z1}
c_{23} &=\frac{\left[\abs{c_{22}}^2-(c_{11}+c_{13})(c_{11}+c^*_{31})\right]c^*_{12}+
	\left[\abs{c_{22}}^2-(c_{11}+c^*_{31})(c_{31}+c_{33})\right]c^*_{32}}{c_{11}+c^*_{31}-c_{13}-c^*_{33}},\\
\label{conditions-msd-along-z2}
c_{21} &= \frac{-(c^*_{13}+c^*_{33})c_{23}-(c^*_{12}+c_{32})c_{22}}{c_{11}+c_{31}}.
\end{align}
Now, the general state is described by twelve parameters. It is cumbersome to work with all the parameters at a time and analyse the nature of squeezing.  Therefore, we shall take different configurations of coupled states and discuss the nature of squeezing  associated with them.  Let us consider the following configurations of the coupled state.

\begin{figure}
	\begin{center}
		\includegraphics[width=4in]{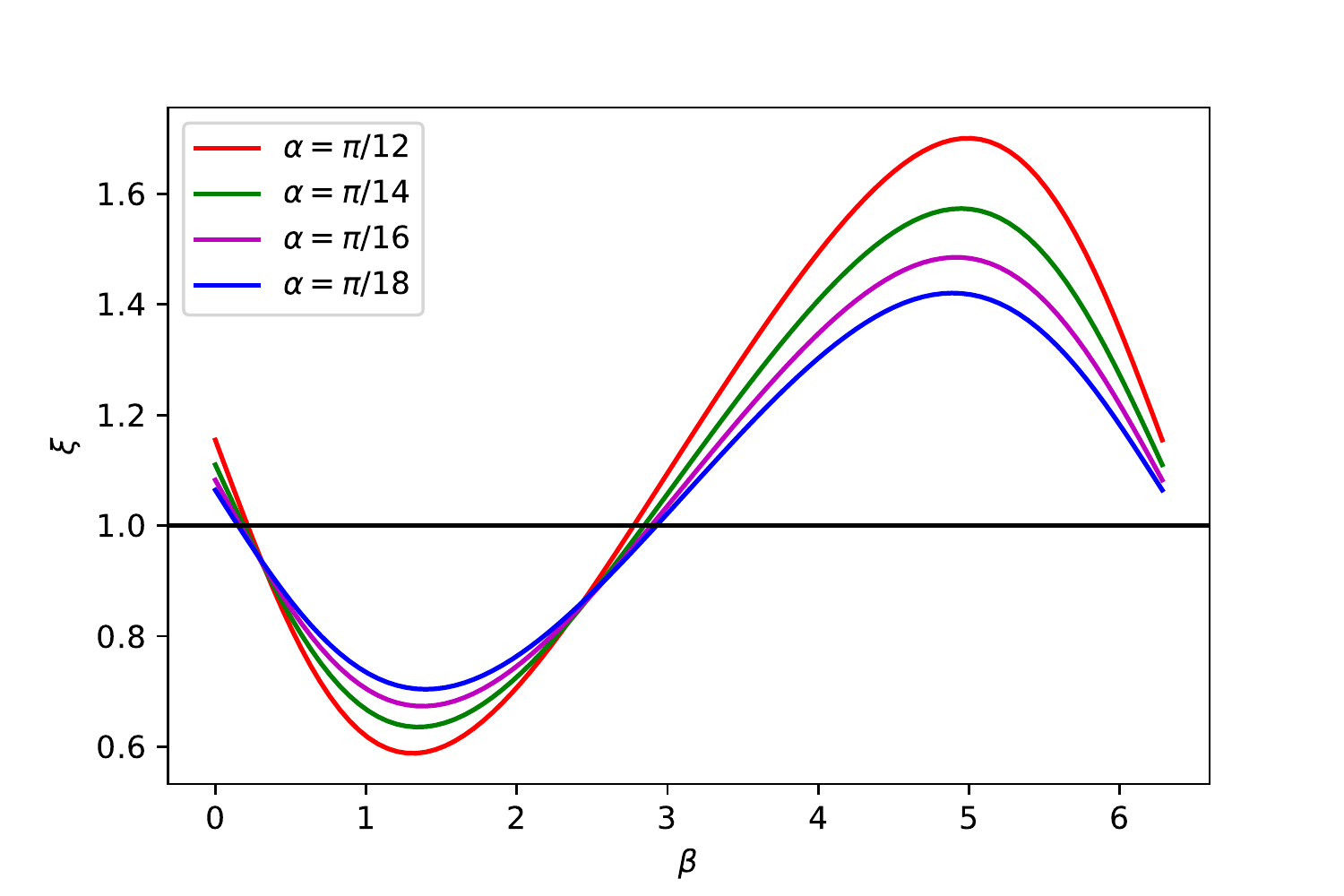}
		\caption{\label{figure4}Variation of squeezing parameter $\xi$ given in eq.(\ref{squeezing-parameter-4}) with respect to $\beta$ for different values of $\alpha$. The plot shows that squeezing exists in {\it configuration 1}.}
	\end{center}
\end{figure}

{\it Configuration 1}: Take only $c_{11}, c_{22}$ and $c_{33}$ as real and nonzero, all others are taken as zero. The general state will be of the form
\begin{align}
\ket{\psi}=\begin{pmatrix}
c_{11} & 0 & 0 & 0 & c_{22} & 0 & 0 & 0 & c_{33}
\end{pmatrix}^T.
\end{align} 
This state satisfies eq.(\ref{condition-for-z-axes}) and eq.(\ref{condition-for-z-axes2}), then mean spin directions are along $z$-axes. For this state, using eq.(\ref{squeezing-parameter}), the squeezing parameter is found to be
\begin{align}
\label{squeezing-parameter-4}
\xi = \frac{c_{11}^2+2c_{22}^2+c_{33}^2-2(c_{11}c_{22}-c_{22}c_{33})}{\abs{c_{11}^2-c_{33}^2}}.
\end{align}
By writing $c_{11}=\sin(\alpha)\cos(\beta),
c_{22}=\sin(\alpha)\sin(\beta),c_{33}= \cos(\beta)$,
we can plot the squeezing parameter and check whether squeezing exists or not. From fig.(\ref{figure4}), we can observe that squeezing exists between 0 and $\pi$ for this configuration of the coupled state. 

\begin{figure}[h]
	\begin{center}
		\includegraphics[width=4in]{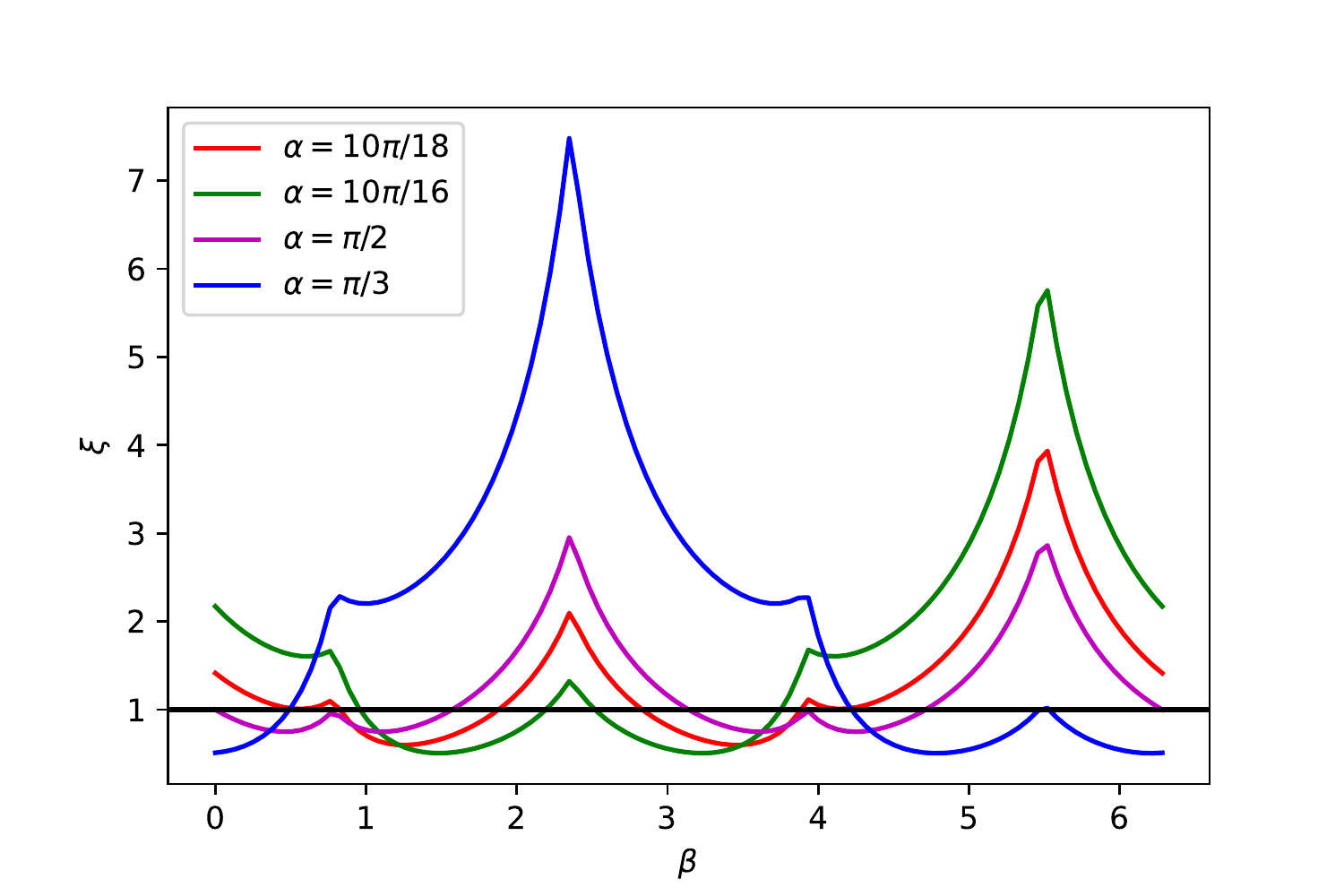}
		\caption{\label{figure5}Variation of squeezing parameter $\xi$ given in eq.(\ref{squeezing-parameter-5}) with respect to $\beta$ for different values of $\alpha$. The plot shows that squeezing exists in {\it configuration 2}.}
	\end{center}
\end{figure}

{\it Configuration 2}: Take only $c_{11},c_{13}$ and $c_{22}$ as real and nonzero, all others are taken as zero. The state will be of the form 
\begin{align}
\ket{\psi}=\begin{pmatrix}
c_{11} & 0 & c_{13} & 0 & c_{22} & 0 & 0 & 0 & 0
\end{pmatrix}^T.
\end{align}
This state satisfies the conditions for mean spin directions along z-axes (see eq.(\ref{condition-for-z-axes}) and eq.(\ref{condition-for-z-axes2})). The squeezing parameter is found to be
\begin{align}
\label{squeezing-parameter-5}
\xi = \frac{c_{11}^2+c_{13}^2+2c_{22}^2-c_{11}c_{13}+2c_{13}c_{22}-2c_{22}c_{11}}{\abs{c_{11}^2+c_{13}^2}+\abs{c_{11}^2-c_{13}^2}}.
\end{align}
Choosing $c_{11}=\sin(\alpha)\cos(\beta), c_{13}= \sin(\alpha)\sin(\beta), c_{22}= \cos(\beta)$, we have plotted the squeezing parameter to check the existence of squeezing.
From fig.(\ref{figure5}), we can see the presence of squeezing for wide range of values. A different configuration of the coupled state having $c_{13}, c_{22}, c_{31}$ as real and nonzero also exhibits squeezing. More or less, the parameters $c_{11},c_{13},c_{22},c_{31}$ and $c_{33}$ having nonzero values exhibits squeezing nature in the coupled state .

{\it Configuration 3}: Take only $c_{12},c_{21}$ and $c_{23}$ as real and non-zero, all others are taken as zero. The state will be
\begin{align}
\ket{\psi}=\begin{pmatrix}
0 & c_{12} & 0 & c_{21} & 0 & c_{23} & 0 & 0 & 0
\end{pmatrix}^T.
\end{align}
 For this state, the squeezing parameter turns out to be
\begin{align}
\label{squeezing-parameter-6}
\xi = \frac{3c_{12}^2+3c_{21}^2+3c_{23}^2-2c_{21}c_{23}+4c_{12}c_{21}-4c_{12}c_{23}}{\abs{c_{12}}^2+\abs{c_{21}^2-c_{23}^2}}.
\end{align}
\begin{figure}[h]
	\begin{center}
		\includegraphics[width=4in]{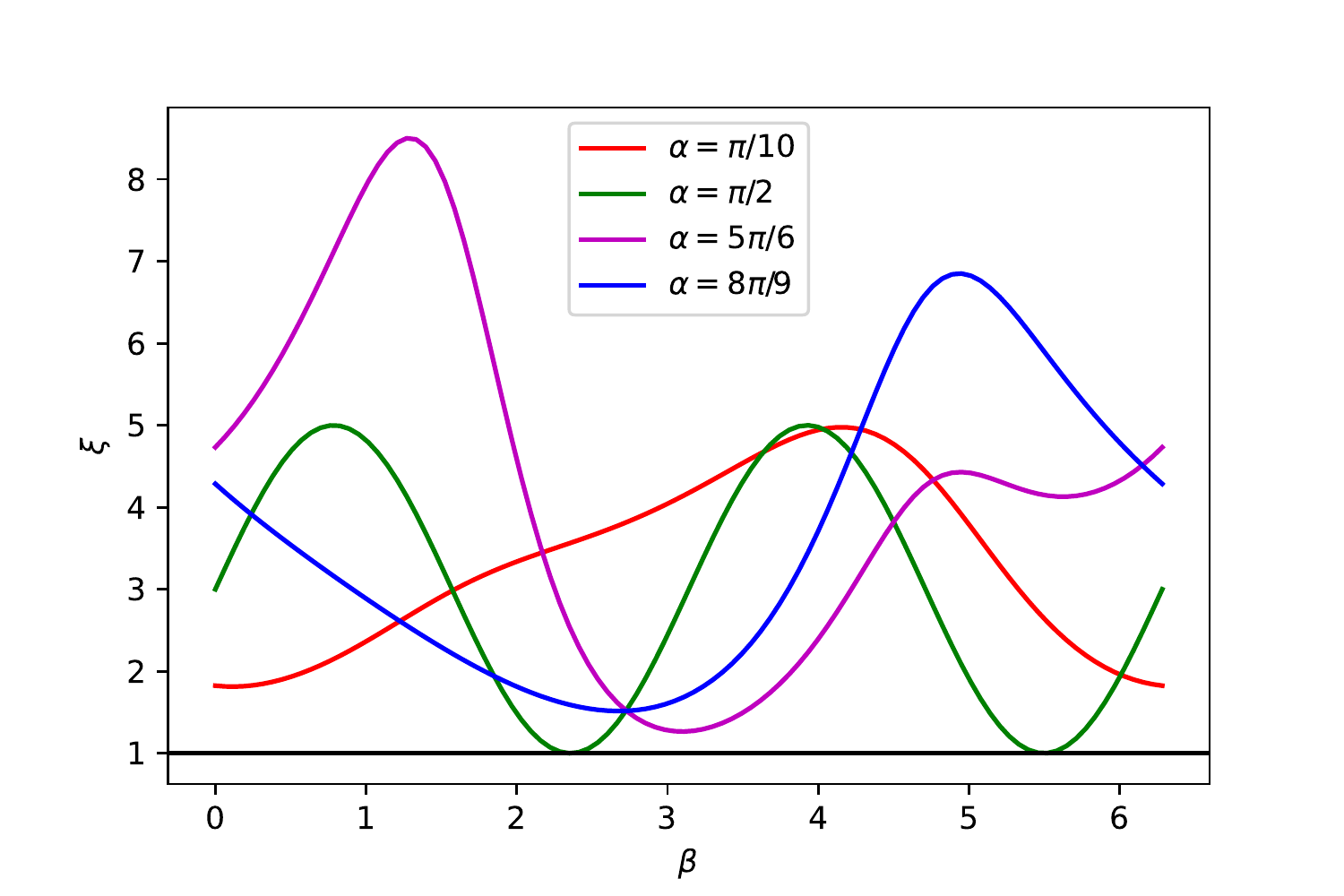}
		\caption{\label{figure6}Variation of squeezing parameter $\xi$ given in eq.(\ref{squeezing-parameter-6}) with respect to $\beta$ for different values of $\alpha$. The plot shows that squeezing does not exist in {\it configuration 3}.}
	\end{center}
\end{figure}
\begin{figure}[h]
	\begin{center}
		\includegraphics[width=4in]{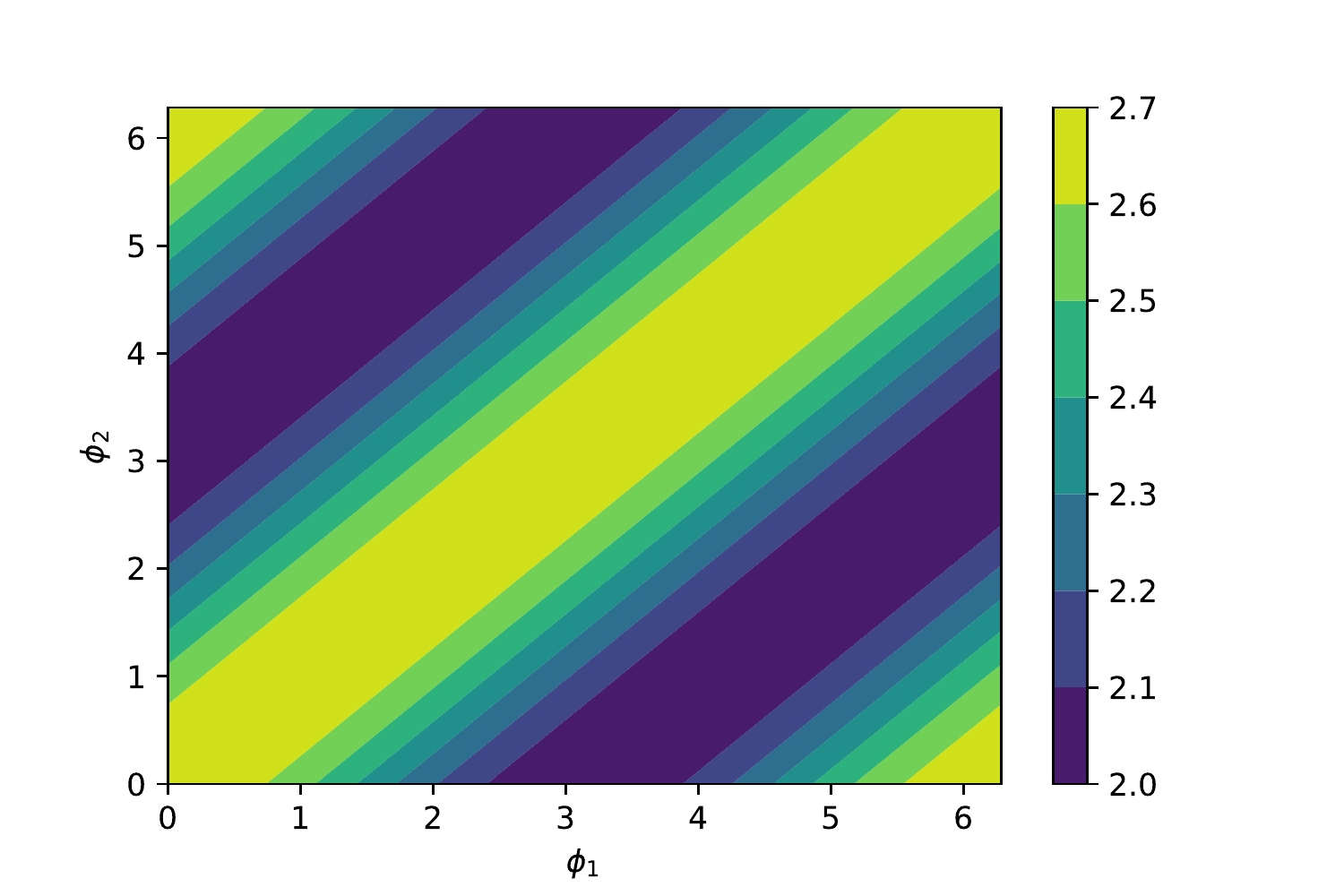}
		\caption{\label{figure7}Variation of squeezing parameter $\xi$ given in eq.(\ref{squeezing-parameter-6}) with respect to $\phi_1$ and $\phi_2$ for $\alpha=\pi/6$ and $\beta=3\pi/4$. The plot shows that squeezing does not exist in {\it configuration 3}.}
	\end{center}
\end{figure}
Choosing $c_{12}=\cos\alpha,c_{21}=\sin\alpha\cos\beta, c_{23}=\sin\alpha\sin\beta$, we plot the squeezing parameter to check for the existence of squeezing. From fig.(\ref{figure6}), we can see that this state does not show squeezing behavior for any values of the parameters. Even by choosing $c_{12}=\cos\alpha,c_{21}=\sin\alpha\cos\beta e^{i\phi_1}, c_{23}=\sin\alpha\sin\beta e^{i\phi_2}$ as complex quantities, the state does not exhibit squeezing(see fig.(\ref{figure7})). This can be seen for all the states having only $c_{12},c_{21},c_{23},c_{32}$ as non zero components. Hence the parameters  $c_{11},c_{13},c_{22},c_{31}$ and $c_{33}$ play an important role in the existence of squeezing.

\section{Generation of squeezing}\label{sec:generation}
So far, we have studied the nature of squeezing for different configurations of the coupled state of two spin-1 systems. In this section, we try to generate coupled squeezed states by external interactions. We know that a Hamilonian consisting only spin operators in linear form can only rotate the state or the coordinate system. Therefore, it is logical to consider the Hamiltonian at least contains spin operators in quadratic form. 

Let us take an initial state of the coupled state to be
\begin{align}
\ket{\psi(0)}=\begin{pmatrix}
1\\
0\\
0
\end{pmatrix}\otimes \begin{pmatrix}
1\\
0\\
0
\end{pmatrix},
\end{align}
which is a coherent state. Consider a Hamiltonian of the form
\begin{align}
H = i\eta [S_{1+}S_{2+}-S_{1-}S_{2-}],
\end{align}
where $S_{i\pm}=S_{ix}\pm i S_{iy}$, $\eta$ is the parameter. The state evolves with time as
\begin{align}
\label{state-1}
\ket{\psi(t)}=\text{exp}[\eta t(S_{1+}S_{2+}-S_{1-}S_{2-})]\ket{\psi(0)}
\end{align}
\begin{figure}[hbt!]
	\begin{center}
		\includegraphics[width=4in]{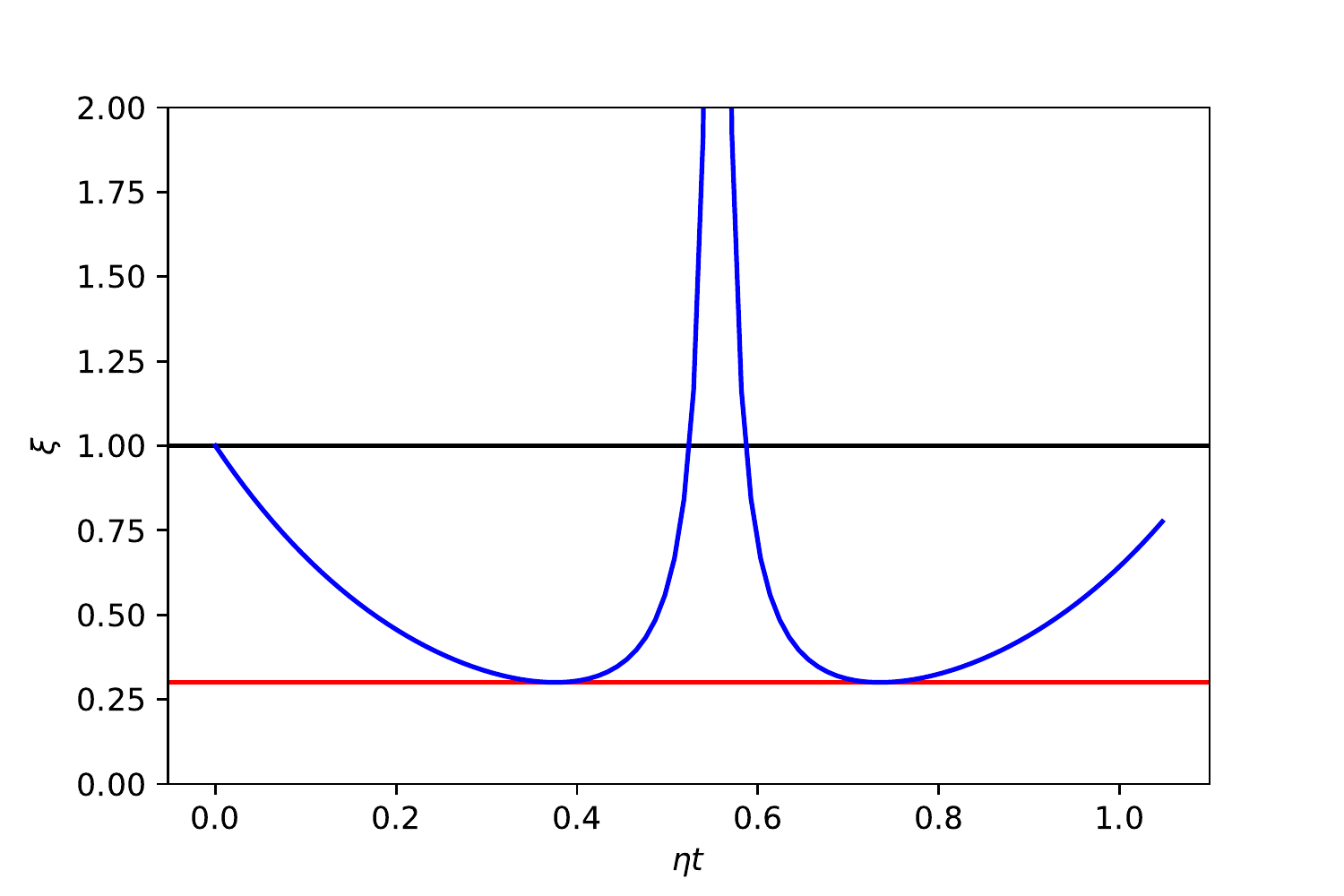}
		\caption{\label{figure8}Plot of squeezing parameter $\xi$ for the state given in eq.(\ref{state-1}) with respect to $\eta t$.}
	\end{center}
\end{figure}
\begin{figure}[hbt!]
	\begin{center}
		\includegraphics[width=4in]{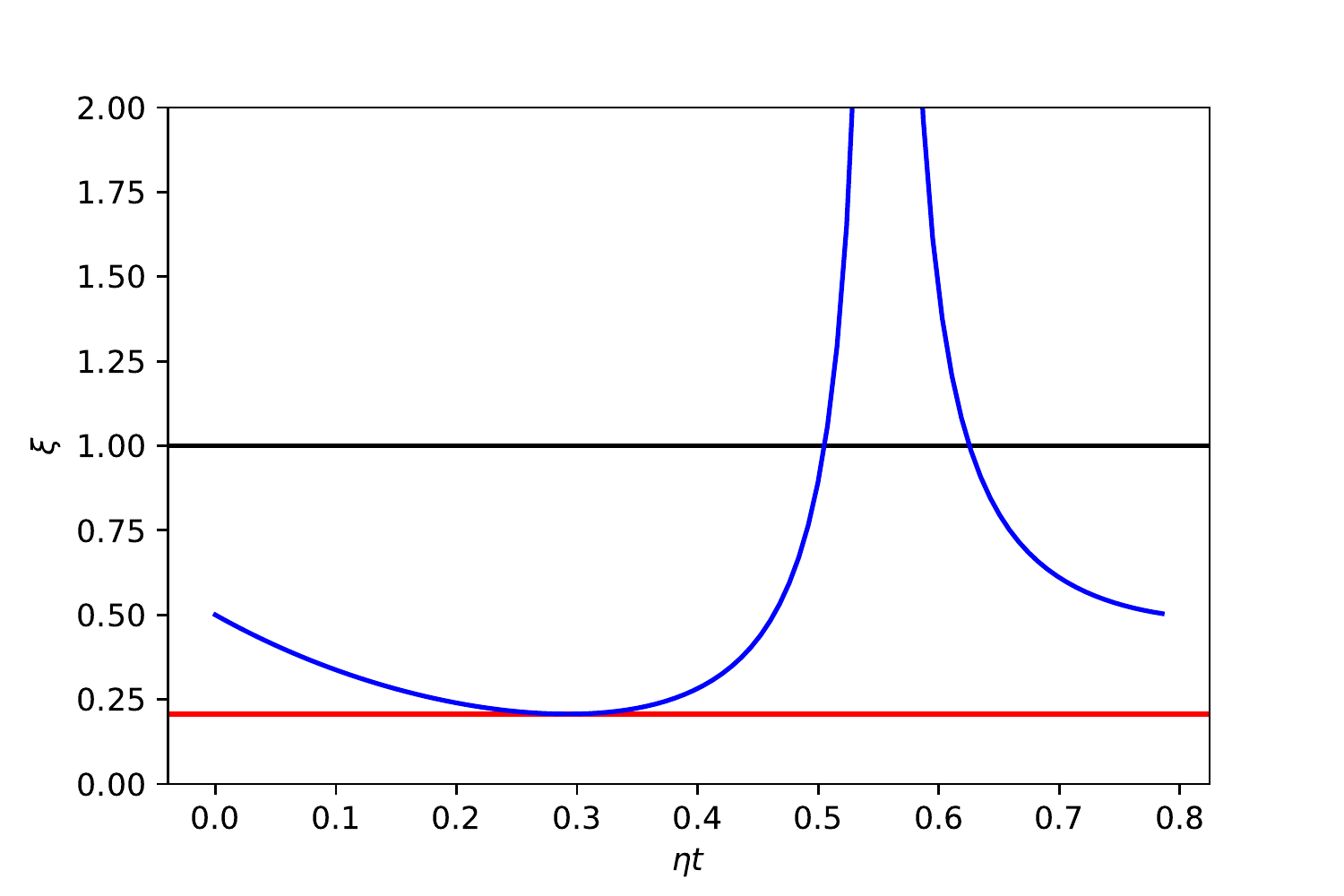}
		\caption{\label{figure9}Plot of squeezing parameter $\xi$ for the state after another interaction $H = \eta[S_{1x}^2S_{2y}^2]$ with respect to $\eta t$.}
	\end{center}
\end{figure}
\begin{figure}[hbt!]
	\begin{center}
		\includegraphics[width=4in]{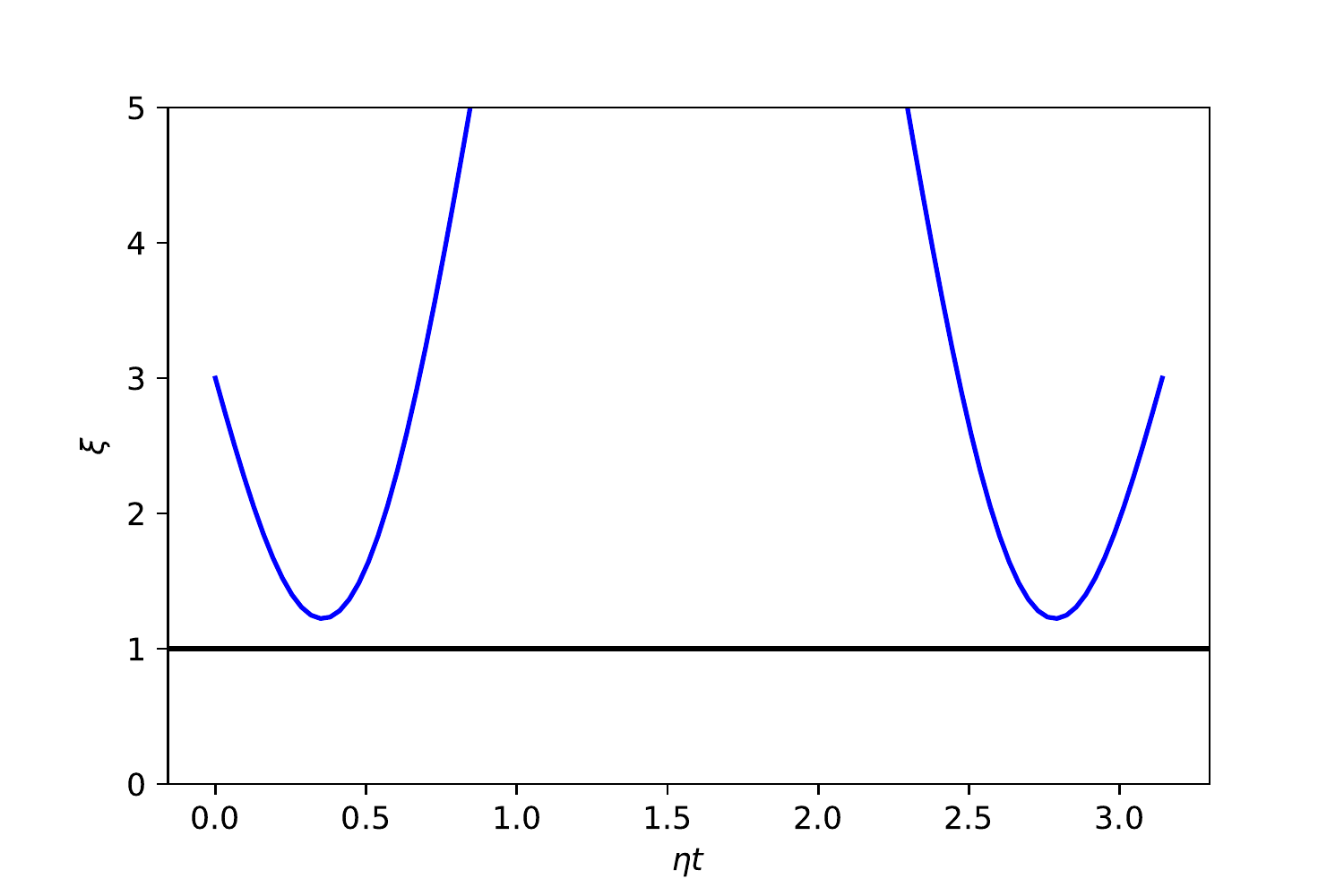}
		\caption{\label{figure10}Plot of squeezing parameter $\xi$ for a different initial state given in eq.(\ref{initial state}) with respect to $\eta t$.}
	\end{center}
\end{figure}
By using squeezing criterion eq.(\ref{squeezing-parameter}), we can investigate the nature of squeezing of the evolved state. From fig.(\ref{figure8}), we can see that the state shows squeezing for wide range of values of the parameter. 
The squeezing parameter can reach below 0.4. One can increase the amount of squeezing by the action of another Hamiltonian. Take $H' = \eta[S_{1x}^2S_{2y}^2]$, act it on the $\ket{\psi(t)}$, then the squeezing parameter can reach 0.2 see fig.(\ref{figure9}). These generated squeezed states have only  $c_{11},c_{13},c_{22},c_{31}$ and $c_{33}$ as non-zero components. This shows that the parameters  $c_{11},c_{13},c_{22},c_{31}$ and $c_{33}$ play a key role in the presence of squeezing in the coupled state. By taking a different initial state as
\begin{align}
\label{initial state}
\ket{\psi(0)}=\begin{pmatrix}
1\\
0\\
0
\end{pmatrix}\otimes \begin{pmatrix}
0\\
1\\
0
\end{pmatrix},
\end{align}
the evolved state does not show squeezing behavior. This state belongs to a class of coupled states, which have only $c_{12},c_{21},c_{23}$, and $c_{32}$ as non zero components,  that does not exhibit squeezing (see fig.(\ref{figure10})).

\section{Conclusion}\label{sec:conclusion}

We have studied the squeezing behaviour of a pure coupled state made up of two spin-1 systems. We have shown that the direct product of two spin-1 systems show squeezing unlike direct product of two spinors. This is due to the existence of self correlation among the two spinors in each spin-1 system. We have also studied the squeezing behaviour of a coupled state which cannot be written as direct product of two spin-1 states. The existence of squeezing is due to both the self correlation among the spinors and the mutual correlation between the spin-1 systems. We have discussed the ways of generating squeezed states and non-squeezed states by taking non-linear spin operators in the Hamiltonian.

\bibliographystyle{alpha}

\end{document}